\documentclass[aps,prl,twocolumn,reprint,showpacs,superscriptaddress,longbibliography]{revtex4-1}
\usepackage{amsmath,amssymb,graphicx,color}
\usepackage{times}
\usepackage{latexsym}
\usepackage{stmaryrd}
\usepackage[table]{xcolor}
\usepackage[colorlinks,allcolors=blue]{hyperref}
\usepackage{amsfonts}
\usepackage{natbib}
\usepackage{appendix}
\usepackage{dsfont}
\usepackage{braket}
\usepackage{scalerel}

\begin{document}
\title{Anomalous criticality with bounded fluctuations and long-range frustration \\induced by broken time-reversal symmetry}
\author{Jinchen Zhao}
\affiliation{Division of Natural and Applied Sciences, Duke Kunshan University, Kunshan, Jiangsu, 215300 China}
\author{Myung-Joong Hwang}
\email{myungjoong.hwang@duke.edu}
\affiliation{Division of Natural and Applied Sciences, Duke Kunshan University, Kunshan, Jiangsu, 215300 China}
\affiliation{Zu Chongzhi Center for Mathematics and Computational Science, Duke Kunshan University, Kunshan, Jiangsu, 215300 China}

\begin{abstract}
We consider a one-dimensional Dicke lattice with complex photon hopping amplitudes and investigate the influence of time-reversal symmetry breaking due to synthetic magnetic fields. We show that, by tuning the total flux threading the lattice with a periodic boundary condition, the universality class of superradiant phase transition (SPT) changes from that of the mean-field fully-connected systems to one that features anomalous critical phenomena. The anomalous SPT exhibits a closing of the energy gap with different critical exponents on both sides of transition and a discontinuity of correlations and fluctuation despite it being a second-order phase transition. In the anomalous normal phase, we find that a non-mean-field critical exponent for the closing energy gap and non-divergent fluctuations and correlations appear, which we attribute to the asymmetric dispersion relation. Moreover, we show that the nearest neighborhood complex hopping induces effective long-range interactions for position quadratures of the cavity fields, whose competition leads to a series of first-order phase transitions among superradiant phases with varying degrees of frustration. The resulting multicritical points also show anomalous features such as two coexisting critical scalings on both sides of the transition. Our work shows that the interplay between the broken time-reversal symmetry and frustration on bosonic lattice systems can give rise to anomalous critical phenomena that have no counter-part in fermionic, spin or time-reversal symmetric quantum optical systems. 
\end{abstract}

\maketitle

{\it Introduction.---} Inspired by the discovery of remarkable phenomena for charged particles moving in magnetic fields such as integer and fractional quantum Hall effects~\cite{tsui_two-dimensional_1982, laughlin_anomalous_1983}, there have been intense theoretical and experimental efforts to realize synthetic magnetic fields for uncharged particles such as photons \cite{umucalilar_artificial_2011,schine_synthetic_2016,roushan_chiral_2017,mann_tunable_2020,Koch:2010eu}, phonons \cite{bermudez_synthetic_2011,abbaszadeh_sonic_2017} and neutral atoms \cite{lin_synthetic_2009,cooper_topological_2019}. In a lattice of photonic resonators, for example, the synthetic magnetic fields have been realized to observe unique topological photonic properties and robust edge states \cite{lu_topological_2014,khanikaev_two-dimensional_2017,ozawa_topological_2019}. Moreover, the chiral photon current due to the time-reversal symmetry breaking induced by synthetic magnetic fields has also been observed \cite{roushan_chiral_2017}. The light-matter interaction between such chiral photons and quantum emitters may give rise to novel quantum optical phenomena \cite{de_bernardis_light-matter_2021,yao_topologically_2013,poshakinskiy_quantum_2021,wang_tunable_2021,Owens.2022}.

A bosonic mode coupled to two-level systems, described by the Dicke model, exhibits a superradiant phase transition (SPT)~\cite{hepp_superradiant_1973,wang_phase_1973,emary_chaos_2003, emary_quantum_2003,nagy_dicke-model_2010,nataf_vacuum_2010,viehmann_superradiant_2011, Kirton.2019} when the spin-boson coupling strength exceeds a threshold. The SPT of the Dicke model belongs to the universality class of fully-connected systems characterized by the mean-field exponents \cite{botet_size_1982,botet_large-size_1983,dusuel_finite-size_2004,emary_quantum_2003,hwang_quantum_2015}; thus, we refer to it as mean-field SPT. Finding SPTs that exhibit critical phenomena that don't belong to this universality class may lead to a discovery of novel phases of coupled light and matter; recently discovered examples include spin glass phases induced by the multimode cavity fields \cite{Gopalakrishnan.2009,strack_dicke_2011,gopalakrishnan_frustration_2011,Gopalakrishnan.2012,Buchhold.2013,Marsh.2021,Chiocchetta.2021,Kelly.2021} and a frustrated SPT in the Dicke lattice~\cite{zhao_frustrated_2022}. Also, a tricritical SPT in the Rabi lattice in the synthetic magnetic field has been discovered~\cite{zhang_quantum_2021,fallas_padilla_understanding_2022}.

In this Letter, we investigate a one-dimensional Dicke lattice model with complex photon hopping amplitudes, whose phase determines the magnetic flux threading the lattice under periodic boundary conditions. We discover that there exists a multicritical magnetic flux point $\theta_c$, above which a mean-field SPT occurs~\cite{Lambert.2003,hwang_quantum_2015,Kirton.2019} and below which an \emph{anomalous} SPT occurs with unusual critical properties that do not belong to the universality class of the fully-connected systems. The anomalous SPT features an anomalous normal phase (NP) in which the fluctuation and correlation do not diverge at the critical point, which shatters the common belief that they always diverge at the critical point. Moreover, the critical exponent of the closing energy gap abruptly changes from a mean-field exponent $1/2$ above $\theta_c$ (NP) to $1$ below $\theta_c$ (anomalous NP); at $\theta_c$, both critical exponents coexist. We show that the anomalous NP emerges when the critical mode acquires a finite momentum, which has an asymmetric dispersion relation due to the time-reversal symmetry breaking. The broken symmetry phase of the anomalous SPT, on the other hand, exhibits a diverging correlation with the critical exponent that are different from that of anomalous NP. Furthermore, we construct an effective semiclassical model for the position quadratures with long-range effective photon hopping interactions. Our effective theory shows that the first-order phase transition from the mean-field to the anomalous superradiant phase occurs when the position quadrature of cavity coherences exhibits frustration. The configuration of order parameters is determined by the relative signs of the effective nearest and the next-nearest interaction, analogous to the $J_1$-$J_2$ Ising model, and therefore a series of first-order phase transitions and multicritical points may emerge as the magnetic flux modulates the sign of both interactions.

\begin{figure*}[t]
	\raisebox{0.\height}{
	\hstretch{1.05}{\includegraphics[width=0.45\textwidth]{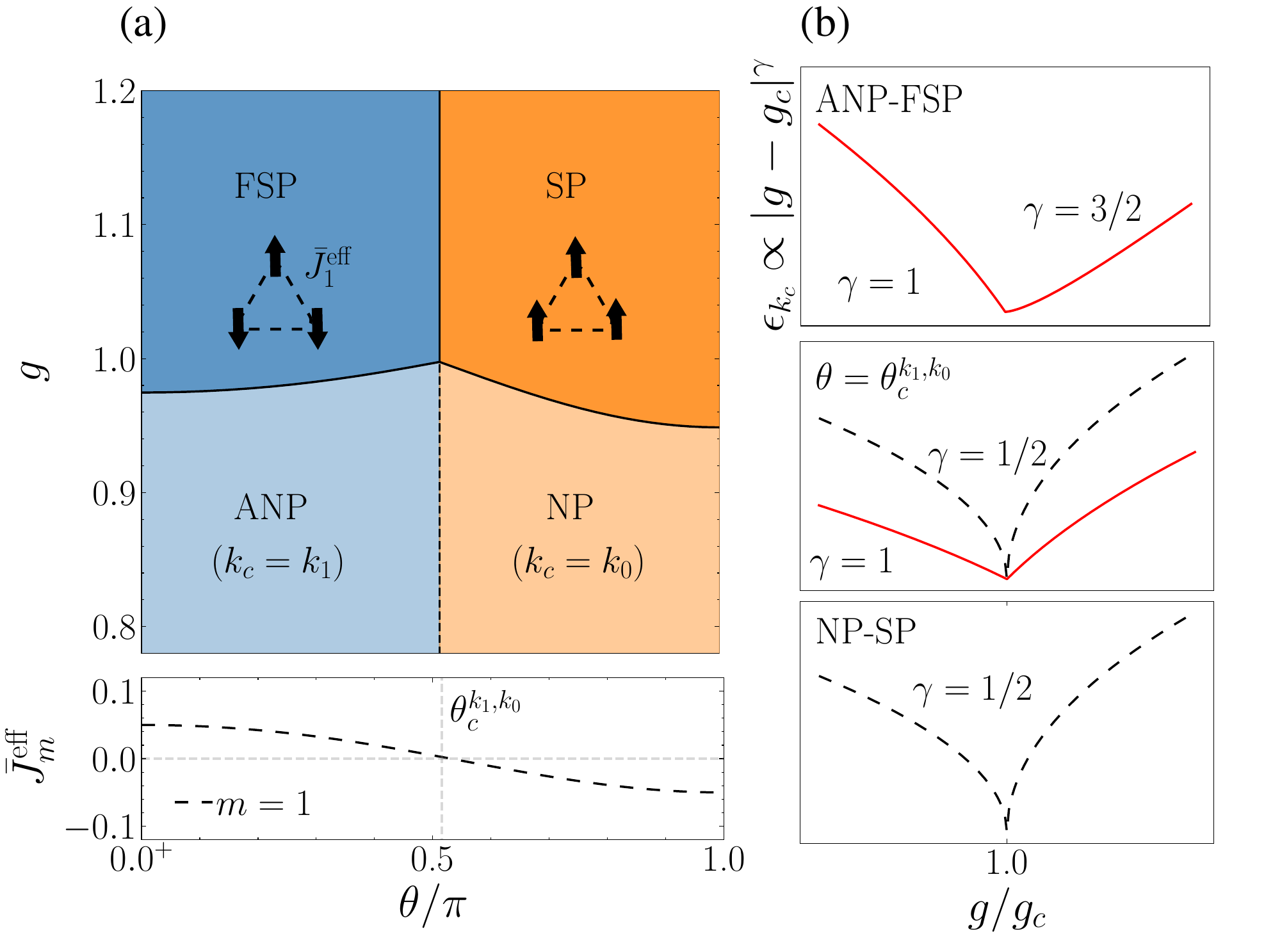}}} \ \ \ \
	\raisebox{0.0\height}{
	\hstretch{1.05}{\includegraphics[width=0.45\textwidth]{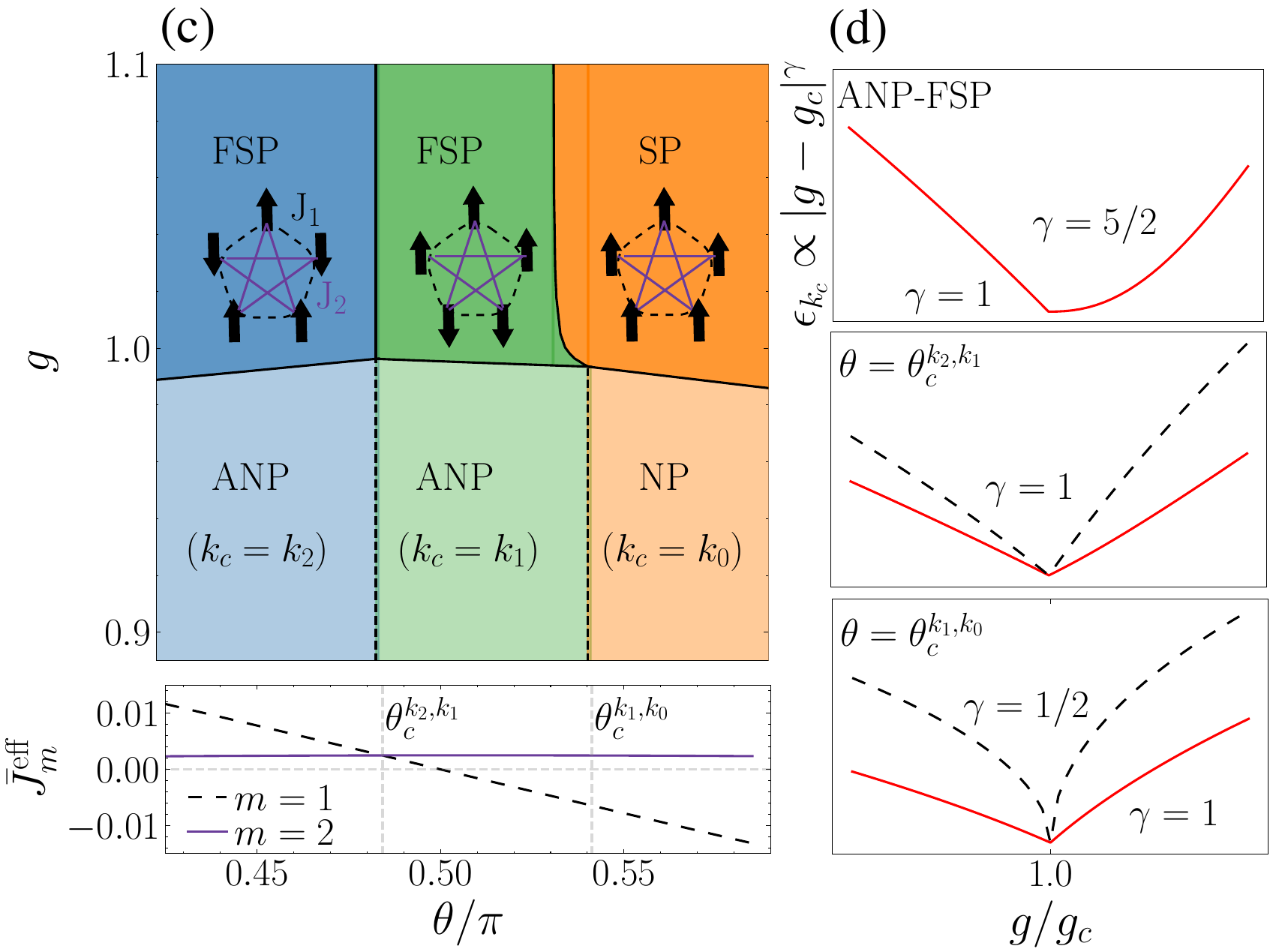}}}
	\caption{Phase diagrams and excitation energies for (a)(b) $N=3$ and (c)(d) $N=5$. (a) and (c) Top panel: Phase diagram in the $g-\theta$ space. Here, NP and SP represent the normal and superradiant phase, respectively, while ANP and FSP stand for anomalous NP and frustrated SP, respectively. $k_c$ is the momentum of the critical mode in NP, and $k_j=-2\pi j/N$. The arrows denote the sign configurations of the mean values $x_n=\operatorname{Re}(\braket{a_n})$, where up arrows denote $x_n>0$, and down arrows denote $x_n<0$. The lines connecting lattice sites correspond to the effective hopping shown in the bottom panel. (a) and (c) Bottom panel: The effective photon hopping $\bar J_m^{\mathrm{eff}}$ in the mean-field energy as a function of $\theta$. Here, $\bar J_1^{\mathrm{eff}}$ is the nearest-neighbor hopping, while $\bar J_2^{\mathrm{eff}}$ is the next-nearest-neighbor hopping. (b) and (d) Critical excitation energies as a function of $g$ for different values of $\theta$. Here, $\gamma$ denotes the scaling of the corresponding excitation, and the upper-right labels denote the region the SPT belongs to. ANP-FSP corresponds to $0<\theta< \theta_c^{k_1,k_0}$, except for the flux critical points $\theta_c^{k_{j+1},k_j}$ (note that the time-reversal symmetric case \cite{zhao_frustrated_2022} is excluded). NP-SP corresponds to $\theta_c^{k_1,k_0}<\theta<\pi$.}
	\label{fig:phase}
\end{figure*}

Our study therefore shows that the presence of the synthetic magnetic field gives rise to the anomalous SPT and anomalous multicritical points with critical properties that has no counter-part in time-reversal symmetric case~\cite{zhao_frustrated_2022} and that are not commonly found in statistical physics systems \cite{cardy_scaling_1996,kardar_statistical_2007,Sachdev:2011uj}; two most important characteristics of the anomalous SPT are the following:  1) critical exponents on both sides of the critical point are different from each other; this adds an experimentally accessible counter-example to the common expectation that critical exponents on both sides of the critical point are same due to the identical renormalization group properties~\cite{leonard_critical_2015}. 2) The fluctuation and correlation are \emph{bounded} at the critical point in the anomalous NP and therefore they become discontinuous across the anomalous SPT, despite it being a second-order phase transition.

{\it Model.---} We consider a Dicke lattice model where each lattice site realizes the Dicke model and neighboring lattices are connected by the photon hopping interaction with complex amplitudes. The model Hamiltonian reads

\begin{eqnarray}
\begin{aligned}
	H_N &= \sum_{n=1}^N \left[H_n+J\left(e^{i\theta}a_n^{\dagger}a_{n+1}+h.c.\right)\right]\label{eq:DickeH}\\
	H_n &=\omega a_{n}^{\dagger} a_{n}+\Omega J_n^z+\frac{2\lambda}{\sqrt{N_a}} \left(a_{n}+a_{n}^{\dagger}\right)J_n^x
\end{aligned}
\end{eqnarray}
with a periodic boundary condition $a_{N+1}=a_{1}$ to form a loop. The phase $\theta\in(0,\pi)$ represents the total flux of synthetic magnetic fields threading the loop. At $n$th lattice site, the oscillator of frequency $\omega$ is described by an annihilation operator $a_n$ and there is an ensemble of $N_a$ spins of frequency $\Omega$ described by collective spin operators $J_n^{x,z}$. $\lambda$ is the local spin-boson coupling strength. The Hamiltonian $H_N$ commutes with the parity operator $\Pi=\exp\left[{i\pi\sum_{n=1}^N (a_n^{\dagger}a_n+J_n^z+\frac{N_a}{2})}\right]$ and thus respects a global $Z_2$ symmetry in addition to the translational symmetry. However, the time reversal symmetry is broken due to the synthetic magnetic field. Note that we consider the limit of an infinite number of atoms in each cavity $N_a\rightarrow\infty$; thus, for any number of lattice sites $N$, the system realizes the thermodynamic limit of infinite particles. Below, we find a rich phase diagram as a function of $g=2\lambda/\sqrt{\omega \Omega}$ and $\theta$ and anomalous critical properties as shown in Fig.~\ref{fig:phase} for $N=3, 5$.

{\it Anomalous normal phase with bounded fluctuation and correlation.---} Let us begin by investigating the normal phase. In the thermodynamic limit, we introduce the Holstein-Primakoff transformation $J_{n}^{+}\simeq\sqrt{N_{a}}b_{n}^{\dagger}$ and $J_{n}^{z}=N_{a} / 2-$ $b_{n}^{\dagger} b_{n}$ with $\left[b_{n}, b_{n}^{\dagger}\right]=1$ and perform a Fourier transform, $a_{n}^{\dagger}=\sum_{k} e^{i kn} a_{k}^{\dagger} / \sqrt{N},b_{n}^{\dagger}=\sum_{k} e^{i kn} b_{k}^{\dagger} / \sqrt{N}$ with $k=0,\pm 2\pi/N, \cdots, \pm (N-1)\pi/N$, to derive the effective Hamiltonian
\begin{equation}\label{eq:DLeffNP}
 H_{\mathrm{np}}=\sum_{k}\left[\omega_k a_{k}^{\dagger}a_{k}+\Omega b_k^{\dagger}b_k-\lambda(a_{k}+a_{-k}^{\dagger})(b_{-k}+b_k^{\dagger})\right],
\end{equation}
where $\omega_k=\omega+2J\cos(\theta-k)$. Note that only the modes with the same magnitude of momentum, $a_{\pm k}$ and $b_{\pm k}$, are coupled with each other; thus, the Hamiltonian can be diagonalized for each $k$. Four excitation energies are given by
\begin{equation}\label{eq:ExcitNP}
\varepsilon_k^{(\pm)}= \sum_{j=1}^2 \sqrt{A^{(\pm)}_{j,k}} + \Delta_k,
\end{equation}
where $\Delta_k=(\omega_k-\omega_{-k})/4$. The superscripts $(\pm)$ denote the upper and lower branches of the excitation spectra, respectively, and the expressions of $A^{(\pm)}_{j,k}$ can be found in \cite{sup}. The upper branch is always gapped, $\varepsilon_k^{(+)}>0$. For each momentum $k$, the lower branch excitation $\varepsilon_k^{(-)}$ can become zero. By solving $\varepsilon_k^{(-)}=0$, we find $g_k(\theta)=\sqrt{2\omega_{k}\omega_{-k}/\omega(\omega_{k}+\omega_{-k})}$. At a given $\theta$, the lowest value of $g_k$ defines the critical point as $H_{\mathrm{np}}$ becomes unstable above this point, indicating the emergence of the superradiant phase. Namely, the critical point at $\theta$ is
\begin{equation}
\label{eq:cp}
g_c(\theta)= \min\{g_{k_j}(\theta)|k_j\},
\end{equation}
where $k_j=-2\pi j/N,~j=0,1, \cdots, (N-1)/2$. We denote the momentum mode that realizes the minimum in Eq.~\ref{eq:cp} as $k_c$, the critical momentum. As the magnetic flux $\theta$ is varied, the critical momentum $k_c$ also changes; thus, we define the flux critical points $\theta_c^{k_{j+1},k_j}$, which mark the boundary between the regions where the modes $k_{j+1}$ and $k_j$ become critical, respectively. These points are found by solving $g_{k_{j+1}}=g_{k_j}$ for $\theta$. We find that $0<\theta_c^{k_{(N-1)/2},k_{(N-3)/2}}<\theta_c^{k_{(N-3)/2},k_{(N-5)/2}}<\cdots<\theta_c^{k_1,k_0}<\pi$. When $\theta>\theta_c^{k_1,k_0}$, the zero-momentum mode becomes critical, i.e., $k_c=k_0$, with a mean-field exponent of $1/2$. Interestingly, for $\theta<\theta_c^{k_1,k_0}$, a non-zero momentum mode becomes critical, i.e., $k_c\neq k_0$, with a non-mean-field exponent of $1$ [see Fig.~\ref{fig:phase} (b) and (d)]. We refer to this as an anomalous NP, which spans $g<g_c(\theta<\theta_c^{k_1,k_0})$.

To understand the emergence of the anomalous NP, we note that $\varepsilon_k$ from Eq. (\ref{eq:ExcitNP}) consists of a sum of square root terms and a constant shift $\Delta_k=(\omega_k-\omega_{-k})/4=J\sin{\theta}\sin{k}$. The latter is the difference in frequencies of lattice photons with opposite momentums and it is non-zero only for $k\neq0$ when the time-reversal symmetry is broken ($\theta\neq0,\pi$). For $k_c=k_0$, $\Delta_{k_c=k_0}=0$; therefore, $\varepsilon_{k_c=k_0}^{(-)}$ closes the gap with the square root  $\varepsilon_{k_c=k_0}^{(-)}\propto|g-g_c|^{1/2}$, a typical mean-field behavior. For $k_c\neq 0$, however, $\varepsilon_{k_c}^{(-)}$ becomes zero before the square root term becomes singular due to the cancelation with $\Delta_{k_c\neq k_0} <0$. In this case, the energy gap closes when an analytical function simply crosses the zero and the exponent becomes $1$, i.e. $\varepsilon_{k_c\neq k_0}^{(-)}\propto|g-g_c|^{1}$. Furthermore, at the boundary between the normal phase and anomalous normal phase, namely, at $g_c(\theta_c^{k_1,k_0})$, we find that both the $k_0$ and $k_1$ mode simultaneously become critical, whose scaling exponents are $1/2$ and $1$. In addition, at other flux critical points between the non-zero momentum modes, i.e., $\theta_c^{k_{i+1},k_i}$ with $i>0$, the two critical excitations with an identical exponent $\gamma=1$ appear [see Fig. \ref{fig:phase}(b)].

\begin{figure}[t]
	\hstretch{1.05}{\includegraphics[width=0.4\textwidth]{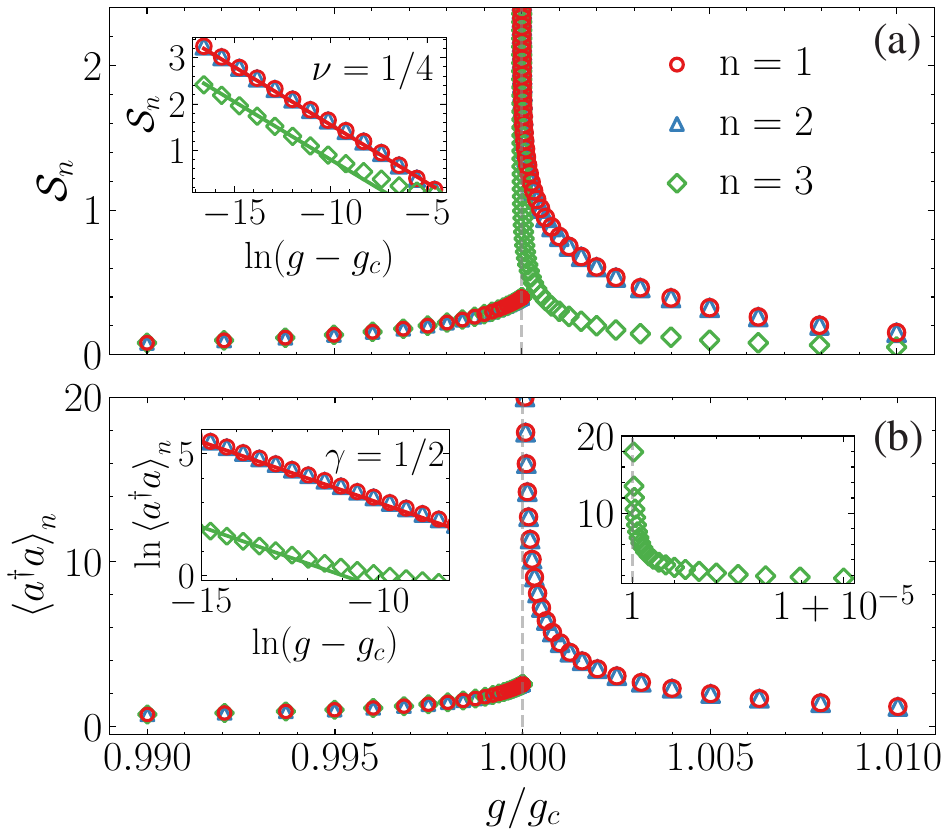}}
	\caption{(a) Bipartite entanglement between $n$th site with the two remaining sites and (b) photon number for $n$th cavity as a function of $g/g_c$ for $N=3$ and $\theta<\theta_c$ across the anomalous SPT. Lines are analytical and shapes are numerical results. Both quantities are bounded in the anomalous NP and divergent in the frustrated SP. The insets on the left side show scaling of the two quantities in the frustrated SP, both of which are $\gamma=1/2$. In the right inset of (a) and (b), the divergence of the $3$rd site is given separately as its peak is much narrower than the rest.}
	\label{fig:entanglephoton}
\end{figure}

In the anomalous NP, the local photon number $\braket{a^{\dagger}a}_n$ and the bipartite entanglement $\mathcal{S}_n$ between the $n$th site and the rest of chain remains finite at the critical point [Fig.~\ref{fig:entanglephoton}(a)]. This is a striking observation because the fluctuation and correlation are typically expected to diverge at the critical point~\cite{Sachdev:2011uj,Osterloh.2002,Lambert.2003}. To gain further insight, we adiabatically eliminate the atomic degrees of freedom in the infinitely frequency ratio limit ($\Omega/\omega\rightarrow\infty$)~\cite{hwang_quantum_2015}, and derive the analytical expression~\cite{sup} for the excitation energy as $\varepsilon_k=\sqrt{A_k}+2\Delta_k$ and the photon number as 
\begin{equation}
	\braket{a_{n}^{\dagger}a_{n}}_{\mathrm{np}}=\frac{1}{N}\sum_k \left[\frac{\omega_{k}+\omega_{-k}}{2\sqrt{A_k}}-\frac{2\sqrt{A_k}}{\omega_{k}+\omega_{-k}}-2\right].
\end{equation}
From this, we see that $\braket{a_{n}^{\dagger}a_{n}}_{\mathrm{np}}\rightarrow\infty$ only if $\sqrt{A_k}\rightarrow 0$ and for the anomalous NP with $\Delta_k<0$, the photon number is bounded. This in turn leads to the bounded entanglement among cavity fields as the entanglement is generated by multimode squeezing with a bounded photon number. On the other hand, for $\theta>\theta_c^{k_1,k_0}$, both $\braket{a^{\dagger}a}_n$ and $\mathcal{S}_n$ diverges with mean-field exponent~\cite{sup} as the photon number is proportional to the inverse of the square root term $\sqrt{A_k}$ in $\varepsilon_k$.

{\it Multicriticality and frustration in the broken symmetry phase.---} When $g>g_c(\theta)$, a second-order continuous phase transition occurs giving rise to spontaneous coherence i.e. $\braket{a_n}=x_n+iy_n\neq 0$. We first replace the operators in Eq.~(\ref{eq:DickeH}) with their mean values to derive the mean-field energy (see Supplementary Material \cite{sup} for the validity of the mean-field approximation). By minimizing the mean-field energy over the atomic degree of freedom~\cite{sup}, we have
\begin{equation}\label{eq:Exy}
\begin{aligned}
	\bar E_N=\sum_{n=1}^N\Big[&\bar x_n^2+\bar y_n^2-\frac{1}{2}\sqrt{1+4g^2\bar x_n^2}
	+2\bar J\cos{\theta}(\bar x_n \bar x_{n+1}\\
	&+\bar y_n \bar y_{n+1})+2\bar J\sin{\theta}(\bar x_{n+1}\bar y_n-\bar x_n \bar y_{n+1})\Big],
\end{aligned}
\end{equation}
where $\bar{J}=J/\omega$, $\bar{E}=E / N_a\Omega$, $\bar{x}_{n}=\sqrt{\omega_{0}/N_a\Omega}~x_n$, $\bar{y}_{n}=\sqrt{\omega_{0}/N_a\Omega}~y_{n}$. We numerically minimize Eq. (\ref{eq:Exy}) and draw the phase diagram for $N=3$ and $N=5$ as shown in Figs. \ref{fig:phase}(a) and \ref{fig:phase}(c). For odd $N$, there exists $(N-1)/2$ first-order transition lines that meet with the continuous transition line $g_c(\theta)$ at each flux critical point $\theta_c^{k_i,k_{i+1}}$, making them tricritical points.

Here, we show that the origin of first-order transitions is the frustration of cavity fields, which leads to the recently discovered frustrated superradiant phase~\cite{zhao_frustrated_2022}. To this end, we derive an effective mean-field energy for the position quadrature $x_n$ of the cavity fields only by eliminating the momentum quadrature $y_n$ at the global minimum of Eq. (\ref{eq:Exy})~\cite{sup}, which leads to
\begin{equation}\label{eq:Ex}
	\bar{E}_N^{\mathrm{GS}}=\sum_{n=1}^{N}\left[\bar x_n^2-\frac{1}{2}\sqrt{1+4g^2\bar x_n^2}+\sum_{m=0}^{(N-1)/2}\bar J^\textrm{eff}_{m} \bar x_n \bar x_{n+m}\right].
\end{equation}
Eq.~(\ref{eq:Ex}) shows that the nearest-neighborhood complex photon hopping effectively realizes long-range interactions among $\bar x_n$ mediated by $\bar y_n$. In particular, we find that the dominant terms are the nearest and next-nearest neighborhood interaction, $|\bar J^\textrm{eff}_{m>2}|\ll |\bar J^\textrm{eff}_{1,2}|$. As the flux modulates signs and magnitudes of $\bar J^\textrm{eff}_{2}$ and $\bar J^\textrm{eff}_{1}$, frustrated sign configurations for $\bar x_n$ may occur, analogous to the $J_1$ and $J_2$ Ising model. We illustrate this point using $N$ odd lattices.

For $N=3$, Eq. (\ref{eq:Ex}) becomes identical with the mean-field energy of the Dicke lattice model with a real photon hopping~\cite{zhao_frustrated_2022}. As shown in Fig.~\ref{fig:phase} (a), $\bar J^\textrm{eff}_1$ changes the sign at the critical flux point $\theta=\theta_c^{k_1,k_0}$. Therefore, the broken symmetry phase undergoes a first-order phase transition between the non-frustrated SP for $\bar J^\textrm{eff}_1<0$ and the frustrated SP for $\bar J^\textrm{eff}_1>0$ with the ground-state degeneracy $D=6$. We note that a similar phase diagram has been found in the Rabi triangle model~\cite{zhang_quantum_2021}, which can also be understood from our effective description.

For $N=5$, when $\theta<\theta_c^{k_2,k_1}$, one has $\bar J_1^\textrm{eff}>\bar J_2^\textrm{eff}>0$ and the nearest neighbors should be anti-aligned to minimize the energy, which is incompatible with a 1D chain with odd $N$. Therefore, a frustrated configuration emerges where a single pair of neighboring sites are aligned, called a ferromagnetic pair, with $D=10$. For $\theta>\theta_c^{k_2,k_1}$, one has $\bar J_2^\textrm{eff}>\bar J_1^\textrm{eff}>0$ that favors the next-nearest neighbors $x_n$ and $x_{n+2}$ to be anti-aligned, which leads to a frustrated configuration with three ferromagnetic pair ($D=10$). This configuration persists even when $\bar J_1^\textrm{eff}$ becomes negative; however, when the negative $\bar J_1^\textrm{eff}$ becomes the dominant energy scale and all $x_n$ have the same sign, leading to a non-frustrated SP with the degeneracy $D=2$. Our analysis can be straightforwardly extended to a larger lattice size for both odd and even $N$~\cite{sup}. For even $N$, if $J_2>0$ is dominant over $J_1$, there could be frustration for odd $N/2$, but no frustration for even $N/2$.


{\it Excitation and fluctuation in the superradiant phase.---} Let us discuss the excitation and fluctuation in various SPs. For detailed derivation, we refer to Ref.~\cite{sup}. For $\theta\geq\theta_c^{k_1,k_0}$, since all mean values $\bar x_n$ are identical with no frustration, the resulting effective Hamiltonian preserves the translational symmetry and we find that the $k_0$ momentum mode becomes critical, with the mean-field exponent $\gamma=1/2$~\cite{sup}. For $0<\theta<\theta_c^{k_1,k_0}$, the translational symmetry of the system is broken due to the frustration. Therefore, we numerically calculate the excitation spectra for $N=3$, $N=5$ [See Fig. \ref{fig:phase}(b)]. For $N=3$, through asymptotic expansions, we analytically derive that the excitation energy gap closes with an exponent $3/2$~\cite{sup}, which agrees with the numerical result. For $N=5$, we find the exponent to be $5/2$. Therefore, we have
\begin{equation}
	\begin{aligned}
		\varepsilon &\propto (g_c-g)^{\gamma_-} &\quad(\textrm{for}~ g<g_c)\\
		\varepsilon &\propto (g-g_c)^{\gamma_+(N)}&\quad(\textrm{for}~ g>g_c)
	\end{aligned}
\end{equation}
with $\gamma_-=1$ and $\gamma_+(N)=N/2$ for $N=3,5$. We note that the possibility of having different critical exponents on both sides of phase transition has been recently discussed in Ref. \cite{leonard_critical_2015} and the anomalous SPT in the synthetic magnetic fields exhibits such unique properties. This is qualitatively different from the scaling behavior of the previously reported frustrated SPT \cite{zhao_frustrated_2022}, where both sides of the transition share a mean-field exponent and an additional non-mean-field scaling appears in FSP. We also calculate the photon number and bipartite entanglement $S_n$ in the SP. Unlike the anomalous NP where both are non-divergent, we find that they do diverge at the critical point as shown in Fig. \ref{fig:entanglephoton} for $N=3$. Therefore, there is a \emph{discontinuity} of both quantities at the critical point of a continuous phase transition. We discovered that the SPT for $0<\theta<\theta_c^{k_1,k_0}$ exhibits highly unusual anomalous critical properties summarized above, and hence we call it an anomalous SPT.

{\it Anomalous multicritical points.---} Finally, we discuss the properties at the multicritical points. We have found that there are two types of multicritical points: i) one is $g_c\big(\theta_c^{k_1,k_0}\big)$, where the boundary between the NP and anomalous NP and the boundary between non-frustrated and frustrated SP meet. At this point, two critical scalings for the closing energy gap with $\gamma=1$ and $\gamma=1/2$ coexist. ii) Others are $g_c\big(\theta_c^{k_{i+1},k_i}\big)$ with $1\leq i \leq (N-3)/2$ where the momentum of the critical mode changes from $k_i$ to $k_{i+1}$ in the normal phase and the sign configuration for the frustrated SP changes. At this point, there are two critical modes on both sides of the multicritical point, but their exponents are both $\gamma=1$. While it is generally expected that the critical exponents at the multicritical point are different from that of the continuous phase transition, two coexisting critical scalings are unique properties of multicritical points of the anomalous SPT.

{\it Discussions.---} 
The Dicke lattice model in the synthetic magnetic fields can be realized in various quantum systems that consist of coupled spins and bosons. The local spin-boson interactions can be implemented using ion-traps~\cite{safavi-naini_verification_2018,cohn_bang-bang_2018}, superconducting circuits~\cite{zou_implementation_2014,nataf_vacuum_2010,viehmann_superradiant_2011}, and cavity QED~\cite{baumann_dicke_2010,kroeze_spinor_2018,klinder_dynamical_2015}. Moreover, the photon or phonon hopping can be engineered to form a desired lattice with a complex hopping energy~\cite{roushan_chiral_2017,yang_synthesis_2019,mathew_synthetic_2020,chen_synthetic_2021,li_atom-optically_2022,del_pino_non-hermitian_2022,frolian_realizing_2022}. Our work demonstrates that the breaking of the time-reversal symmetry offers a unique mechanism for the normal phase of lattice bosons to become unstable with bounded fluctuation and that the complex nearest-neighborhood hopping amplitudes effectively mediate long-range interactions which may lead to exotic frustrated quantum phases of coupled light and matter. How the unique scaling properties of the anomalous SPT discovered here could be used for the applications in critical metrology based on quantum optical models like Dicke/Rabi model is an interesting topic for future investigations \cite{garbe_critical_2020,chu_dynamic_2021,ilias_criticality-enhanced_2022}.

{\it Note added.---} Upon completion of this work, we became aware of a recent work on the Rabi lattice models in synthetic magnetic fields~\cite{fallas_padilla_understanding_2022}. 

\emph{Acknowledgments.---}
This work was supported by NSFC under Grant No. 12050410258, the Startup Fund and Summer Research Scholar Program from Duke Kunshan University, Kunshan Municipal Government research funding, and Innovation Program for Quantum Science and Technology 2021ZD0301602.

\end{document}